\documentclass[twocolumn,showpacs,preprintnumbers,fleqn,amsmath,amssymb,aps]{revtex4}

\usepackage{graphicx}
\usepackage{time}
\usepackage{dcolumn}
\usepackage{bm}
\usepackage{amsmath}
\usepackage{amssymb}
\usepackage{amstext}
\usepackage{epsfig}
\usepackage{latexsym}
\usepackage{times}
\usepackage{epstopdf}

\begin{document}

\title{Micromagnetic study of equilibrium states in nano hemispheroidal shells}
\author{Keren Schultz}
\author{Moty Schultz}
\affiliation{Department of Physics, Nano-magnetism Research
Center, Institute of Nanotechnology and Advanced Materials,
Bar-Ilan University, Ramat Gan 5290002, Israel}

\date{\today}

\begin{abstract}

We present results of micromagnetic simulations of thin ferromagnetic nano hemispheroidal shells with sizes ranging from 1 to 50 nm (inside dimensions). Depending on the geometrical and magnetic parameters of the hemispheroidal shell, there exist four different magnetic phases: easy axis, easy plane, onion and vortex. The profile for the vortex magnetization distribution is analyzed and the limitations and applicability of different vortex ansatzes are discussed. In addition, we investigate the total energy density for each of the magnetic distributions as a function of the hemispheroidal shell dimensions.  

\end{abstract}

\pacs{75.10.Hk, 75.40.Mg}

\maketitle

\section {Introduction}

Synthesis and application of magnetic nanoparticles is a fast burgeoning field which has potential to bring significant advance in many fields, ranging from magnetic nanoparticles based cancer therapy to sensors \cite{Wang2013}. The magnetic properties of nanoparticle are important since they enable the manipulation of particles behaviour remotely and therefore provide the means to direct a particles' orientation and translation. Janus particles (JPs) are particles with two dissimilar faces having unique material properties \cite{Perro2005,Walther2008}. JPs provide us with the ability to complete challenging tasks that would be impossible with isotropic particles. Magnetic JPs combine magnetic properties with anisotropy and thus are potential building blocks for complex structures that can be assembled from a particle suspension and can be directed easily by external magnetic fields. JPs with ferromagnetic or paramagnetic properties of various anisotropic shapes have been proposed to have potential application in probing the rheology of complex fluid interfaces \cite{Choi2011}. Magnetic JPs have interesting properties for their application in microfluidics and as nanodelivery systems. In order to control magnetic JPs it is very important to understand their magnetization behaviour.

Masking is the simplest and was one of the first techniques developed for the synthesis of magnetic JPs. In a first step, particles were deposited onto a substrate and then the above-lying surface was coated with a thin shell of a magnetic metal. During the last years there was an increasing interest in studying the magnetic properties of spherical and hemispherical magnetic shells. In all these geometries which were investigated so far, experiment measurements,  analytical calculations and micromagnetic simulations suggest the existence of three magnetic states: uniform state (easy plane), vortex state and onion state \citep{Ulbrich2006,Streubel2012,SHEKA2013,Streubel2012a,Kravchuk2012}. Considerably less attention has been devoted to non spherical shells, such as spheroids.      

In this work, we present micromagnetic simulations of nano hemispheroidal shells, by using Nmag simulator \cite{Fischbacher2007}. The simulations were made for a variety of shells' thicknesses and sizes ranging from 5 to 10 nm and 1 to 50 nm (inside dimensions), respectively. We find the phase diagram of equilibrium magnetization states in these shells. In contrast to the hemispherical magnetic shells which have only three magnetic phases, the hemispheroidal magnetic shells have four magnetic configurations: two phases with homogeneous magnetization (easy axis and easy plane) and two phases inhomogeneous (onion and vortex). The easy axis structure is encountered only in the small and elongated hemispheroidal magnetic shells. Hereafter we consider the vortex magnetization profiles and the limitation and applicability of different vortex ansatzes are discussed. Moreover, we investigate the total energy density for each of the magnetic distributions as a function of the hemispheroidal shell dimensions.

\section {method}

The starting point of magnetization dynamics at zero temperature is from the Landau-Lifshitz-Gilbert (LLG) equation,

\begin{equation}
\label{LLG_eq}
\dfrac{d\textbf{M}}{dt}=-\dfrac{\gamma}{1+\alpha^{2}}[\textbf{M}\times \textbf{H}_{eff}+\dfrac{\alpha}{\vert{\textbf{M}}\vert}\textbf{M}\times(\textbf{M}\times \textbf{H}_{eff})],
\end{equation}

where $\gamma$ is the gyromagnetic ratio and $\alpha$ is the damping constant. There are some publicly available packages for solving LLG equation, like OOMMF \cite{M.DonahueandD.porter1999}, mumax$\rm{^3}$ \cite{Vansteenkiste2014} and NMAG \cite{Fischbacher2007}. 

OOMMF and mumax$\rm{^3}$ are based on discretizing space into small cuboid cells, respectively. One advantage of this method (often called ‘finite differences’) is that the demag field can be computed very efficiently. On the other hand, this method works less well if the geometry shape does not align with a Cartesian grid since the boundary is represented as a staircase pattern.

Nmag's finite elements discretize space into many small tetrahedra. The corresponding approach is based on the Fredkin Koehler hybrid Finite Element/Boundary Element method (FEM/BEM) \cite{Fredkin1990}. The advantage of this method (over $\rm{mumax^3}$ and OOMMF's approach) is that curved and spherical geometries can be spatially resolved much more accurately. However, this method of calculating the demagnetisation field is less efficient than OOMMF’s approach for thin films. In particular, memory requirements for the boundary element method grow as the square of the number of surface points. Note that for simulation of thin films, the hybrid FEM/BEM is likely to require a lot of memory. In order to improve the efficiency of the hybrid FEM/BEM, one can employ techniques which involve some kind of approximation, for example using hierarchical matrices \cite{Knittel2009}. The library Hlib [http://www.hlib.org] contains implementations of this hierarchical matrix methodology. Because our samples have curved geometries we choose Nmag and for the big samples we use Nmag with Hlib.

The cell size of micromagnetic simulations should be equal or smaller than the half size of the smallest feature of interest, so sampling at that length scale is sufficient to capture all relevant details. In our micromagnetic simulations it is the domain wall. For soft materials the domain wall profile is determined by competition between the exchange and demagnetization energies. Therefore, it is common to define magnetostatic exchange length as $l_{ex}=\sqrt{2A/(\mu_0M_s^2)}$ , where $A$ is the exchange constant in J/m and $M_s$ is the saturation magnetization in A/m. In order to get reliable results, in all our simulations the cell size was smaller than half of the exchange length. Good criteria for testing the reliability of the simulations is the maximum angle between two neighbouring cells. Because this is such important issue Nmag, provides this data.         

A spheroid, or ellipsoid of revolution, is a quadric surface obtained by rotating an ellipse about one of its principal axes; in other words, an ellipsoid with two equal semi-diameters. If the ellipse is rotated about its major axis, the result is a prolate (elongated) spheroid. While rotating about its minor axis, the result is an oblate (flattened) spheroid.

The systems studied are hemispheroidal shells with Permalloy parameters: exchange coupling $A=13 \times 10^{-12} \frac{J}{m}$, saturation magnetization $M_s=0.86 \times 10^{6} \frac{A}{m}$ and a megnetocrystalline anisotropy was neglected. Therefore, the exchange length is about 5 nm. We created the shells by rotating a half elliptical ring around the z axis. The internal dimensions of the shells are of a half spheroid with semi-principal axes of length $\rm A_{in}$, $\rm B_{in}$ and $\rm C_{in}$, where $\rm B_{in}=A_{in}$. The outside dimensions of the shells are of a half spheroid with semi-principal axes of length $\rm A_{out}$, $\rm B_{out}$ and $\rm C_{out}$, where $\rm B_{out}=A_{out}$.   

\section {Phase diagram}
 
\begin{figure}
\begin{center}
\includegraphics [scale=0.52]{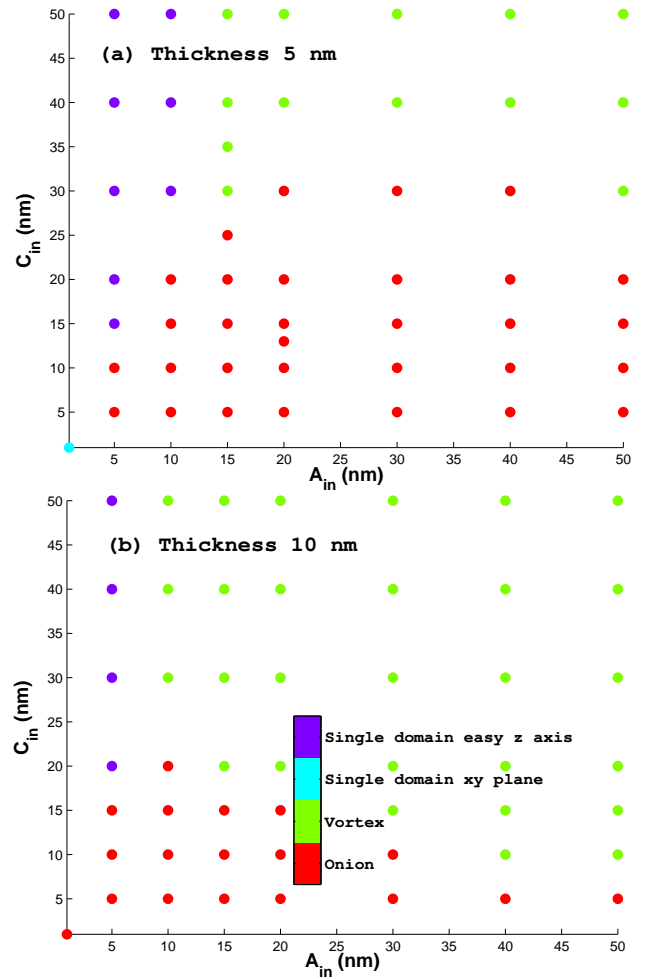}
\end{center}
\caption{Phase diagrams of simulated equilibrium magnetization states in Permalloy hemispheroidal shells. (a) Shell thickness 5 nm. (b) Shell thickness 10 nm.} \label{Phase}
\end{figure}

We simulate the remanent states starting with two different directions of hysteresis loops: along the x direction (in the cutting plane of the hemispheroidal shell) and along the z direction (perpendicular to the cutting plane). Depending on the dimensions of the hemispheroidal shell, there are one or two remanent states. The ground state is the one with the minimum energy. Following such a scheme, we construct phase diagram of equilibrium magnetization structures in the hemispheroidal shell. Since we have three parameters, which define the hemispheroidal shell, $\rm A_{in}$, $\rm C_{in}$  and the thickness of the shell, our phase diagram is a three dimensional. We choose to show the phase diagram with two dimensional plots for clarity (see Figure $\ref{Phase}$). The general properties of the phase diagram are as follows. The ground state of the smallest hemispheroidal shells is the homogeneous easy plane state. As $\rm A_{in}$ and $\rm B_{in}$ increase we switch to onion state and then to vortex state. The vortex state becomes much more preferable while moving to larger thicknesses. In addition, for high aspect ratio ($\rm C_{in}/\rm A_{in}$) the homogeneous easy axis becomes preferable.  

\begin{figure}
\begin{center}
\includegraphics [scale=0.5]{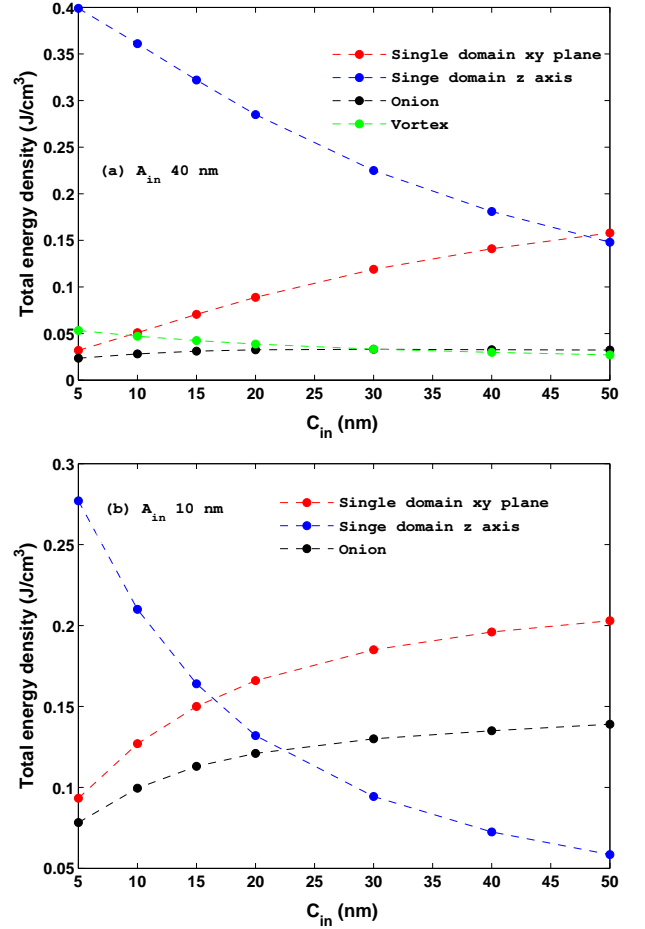}
\end{center}
\caption{Total energies of different magnetization distributions as a function of the semi-principal axis $\rm C_{in}$ for thickness=5 nm and different semi-principal axis (a) $\rm A_{in}$=40 nm (b) $\rm A_{in}$=10 nm . The dashed lines are guides for eyes.} \label{Energy_thickness5}
\end{figure}

The total energy densities were obtained for different hemispheroidal shells' sizes and magnetization distributions, the results of these simulations are presented in Fig $\ref{Energy_thickness5}$. For the onion and vortex states the energies were taken from the zero field data in the hysteresis loops simulations. In order to find the energy density of the single domain z axis and the single domain xy plane in zero magnetic field, much simpler simulations are used; we put the system in the single domain configurations xy plane ($m_x=1$), single domain z axis $(m_z=1)$ and simulate the energy. Figure $\ref{Energy_thickness5}$(a) depicts the total energy densities of wide hemispheroidal shells ($\rm A_{in}$=40 nm) with thickness=5 nm. Due to the magnetostatic energy, the total energy density of the single domain xy plain increases with $\rm C_{in}$. While $\rm C_{in}$ increases, the total energy density of the onion state increases when $\rm C_{in}<$30 nm and decreases when $\rm C_{in}>$ 30 nm; the decrease of the total energy density when $\rm C_{in}>$ 30 nm is very small and therefore unnoticeable in the figure. While $\rm C_{in}$ decreases, the total energy density of the vortex state and the single domain z axis increase. The total energy density of the single domain z axis is very high for all value of $\rm C_{in}$ and decrease with increasing  length of $\rm C_{in}$. Figure $\ref{Energy_thickness5}$(b) shows the total energy densities of narrow hemispheroidal shells ($\rm A_{in}$=10 nm) with thickness=5 nm. As can be seen the energies for the single domain easy axis and the single domain easy plane behave qualitative the same as in Figure $\ref{Energy_thickness5}$(a). There are two main differences between Figure $\ref{Energy_thickness5}$(a) and Figure $\ref{Energy_thickness5}$(b): for $\rm A_{in}$=10 nm the vortex sate is no more a reachable metastable state and the the single domain z axis becomes to be the ground state for $\rm C_{in} >22$ nm.

In the absence of intrinsic anisotropy energy, the reason for all these various ground states is the competition between two energies: the exchange energy and stray field energy. A general solution of the stray field is given by potential theory. The volume magnetostatic charge density $\lambda$ and the surface magnetostatic charge density $\sigma$ are defined in terms of the normalised magnetization: $\lambda=-\nabla\cdot\bold{m}$, $\sigma=\bold{m}\cdot\bold{n}$ where $\bold{n}$ is the outward surface normal. For the vortex state, the surface and the volume magnetostatic charges are almost absent. In the vortex distribution there is a stray field only inside the vortex core.   For the homogeneous magnetization distribution the total energy contains only the contribution of surface magnetostatic charges. These surface magnetostatic charges are the source of demagnetization field which is opposite to the sample magnetization. In general, the demagnetization field along a short axis is stronger than that along a long axis. Due to demagnetization field, the single domain z axis becomes energetically favorable with the particle size decreasing and the aspect ratio $\rm C_{in}/\rm A_{in}$ increasing. For nano particles the single domain configuration becomes the lowest energy state when the particle diameter becomes comparable or smaller than the exchange length. Therefore, the smallest nano hemispheroidal shell ($\rm A_{in}$=1, $\rm B_{in}$=1 and thickness 5 nm) in our simulation has a single domain xy configuration.

 \section{Vortex state}
 
\begin{figure}
\begin{center}
\includegraphics [scale=0.16]{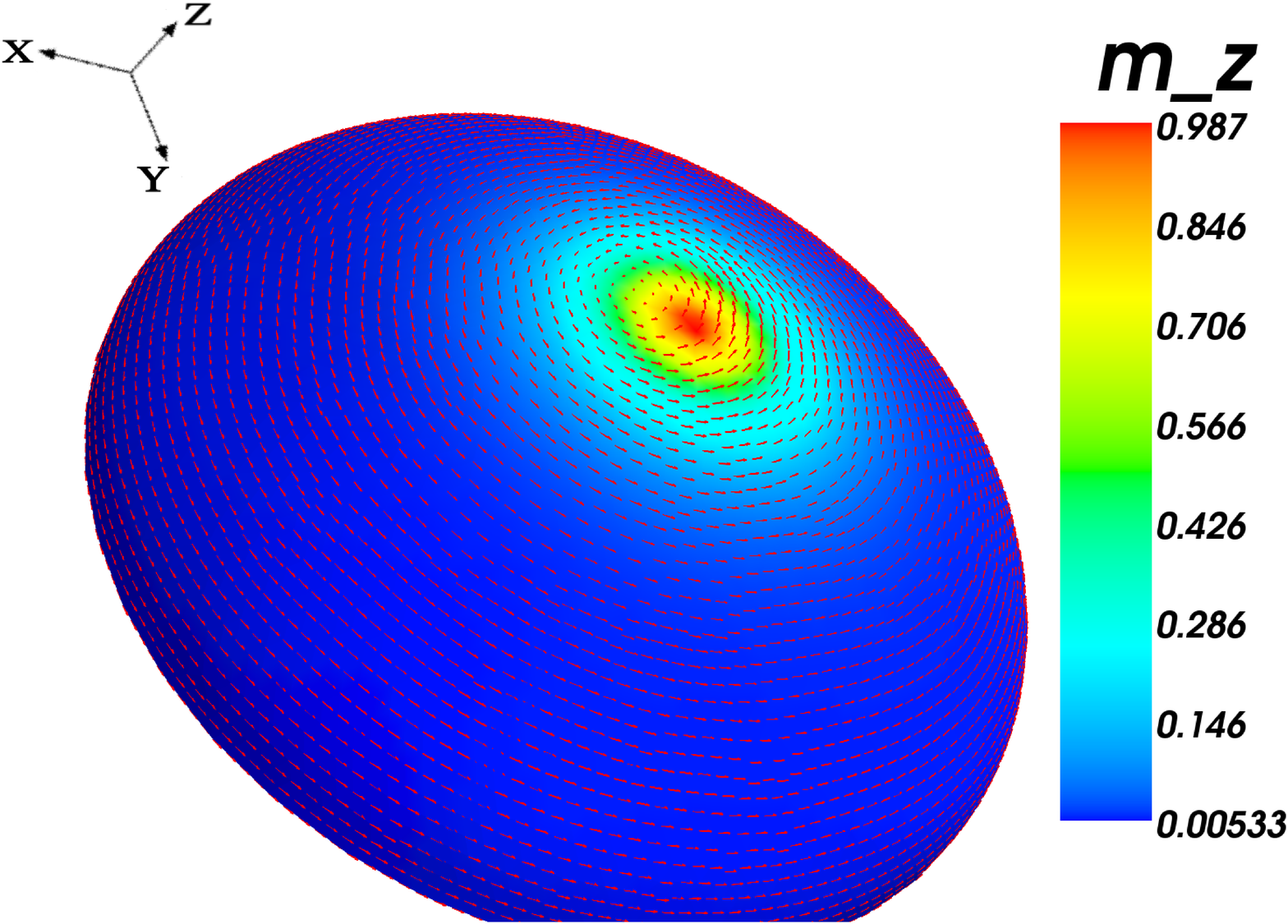}
\end{center}
\caption{The magnetization distribution in vortex state of an oblate hemispheroidal shell with $\rm A_{in}=40$ nm, $\rm C_{in}=20$ nm and thickness 5 nm. The color code represents the normalized z component of the magnetization, emphesizes the core in the center of the vortex.}   \label{oblate_vortex}
\end{figure}

\begin{figure}
\begin{center}
\includegraphics [scale=0.16]{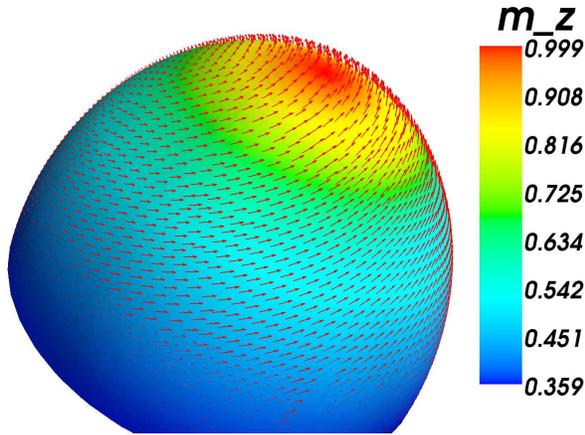}
\end{center}
\caption{The magnetization distribution in vortex state of a prolate hemispheroidal shell with $\rm A_{in}=15$ nm, $\rm C_{in}=20$ nm and thickness 5 nm. The color code represents the normalized z component of the magnetization, shows that the core is not only in the center of the vortex.}   \label{prolate_vortex}
\end{figure}

 The typical vortex magnetization configuration for an oblate hemispheroidal shell and a prolate hemispheroidal shell are depicted in Figures $\ref{oblate_vortex}$ and $\ref{prolate_vortex}$, respectively. While for the oblate hemispheroidal shell we can see clearly a core in the middle of the vortex, in the prolate hemispheroidal shell the core is spread all over the vortex.     
 
 Using the angular parametrisation for the normalised magnetisation 
 \begin{equation}
\label{m_eq}
\bold{m}=\dfrac{\bold{M}}{M}=(\sin\theta \cos\phi, \sin\theta\sin\phi, \cos\theta), 
\end{equation}
 one can describe the vortex solution as follows:
 
\begin{equation}
\label{vortex_eq}
\cos\theta=pf(r),\qquad     \phi=\epsilon\dfrac{\pi}{2}+\chi .
\end{equation}

\begin{figure}
\begin{center}
\includegraphics [scale=0.5]{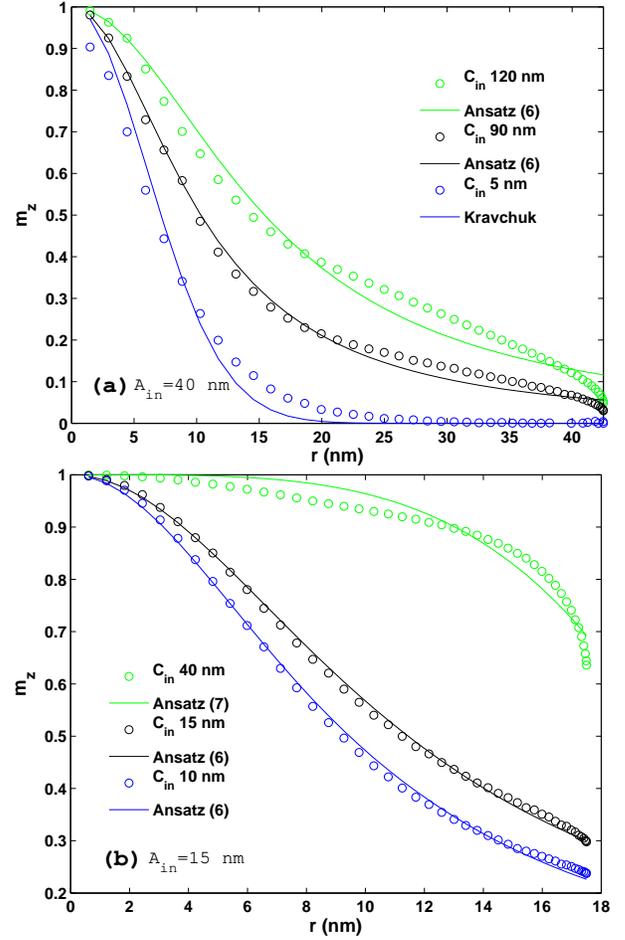}
\end{center}
\caption{The normalized z component of the magnetization as a function of $r$ for hemispheroidal shells (Thickness=5 nm). (a) Symbols correspond to the micromagnetic simulations for semi-principal axis $\rm A_{in}=40$ nm and different $\rm C_{in}$.  (b) Symbols indicate the micromagnetic simulations results for semi-principal axis $\rm A_{in}=15$ nm and three different $\rm C_{in}$ values. Solid lines are the best ansatzes which fit to the simulations' results.}  \label{profile_vortex5}
\end{figure}

 Here, $(r,\chi,z)$ are the cylinder coordinates, $p=\pm1$ is the vortex polarity which describes the vortex core magnetization (up or down) and $\epsilon=\pm1$ is the vortex chirality (clockwise or counterclockwise). The function $f(r)$ describes the out of surface structure of the vortex. In order to find a good fit to the oblate hemispheroidal shell, it is instructive to use an analogy with a vortex profile of the planar disk. There are known several models for describing the vortex in a disk shaped particles. For example the ansatz by Usov and Peschany \cite{N.a1992,Guslienko2004} and the Kravchuk's ansatz \cite{Kravchuk2007} which describes the vortex structure in disks and rings,
\begin{equation}
\label{Kravchuk_eq}
f(r)=e^{-(\dfrac{r}{\xi})^2},
\end{equation}
where the parameter $\xi$ determines the radius of the vortex core. For the hemispherical shells Sheka et al. used a modified version of Usov's and Peschany's model \cite{SHEKA2013} 
 \begin{equation}
\label{Usov_eq}
f(r) = \begin{cases} \dfrac{r_c^2-r^2}{r_c^2+r^2} & \text{if } r<r_c \\ 0 &\text{if } r\geq r_c   \end{cases}.
\end{equation}

\begin{figure}
\begin{center}
\includegraphics [scale=0.5]{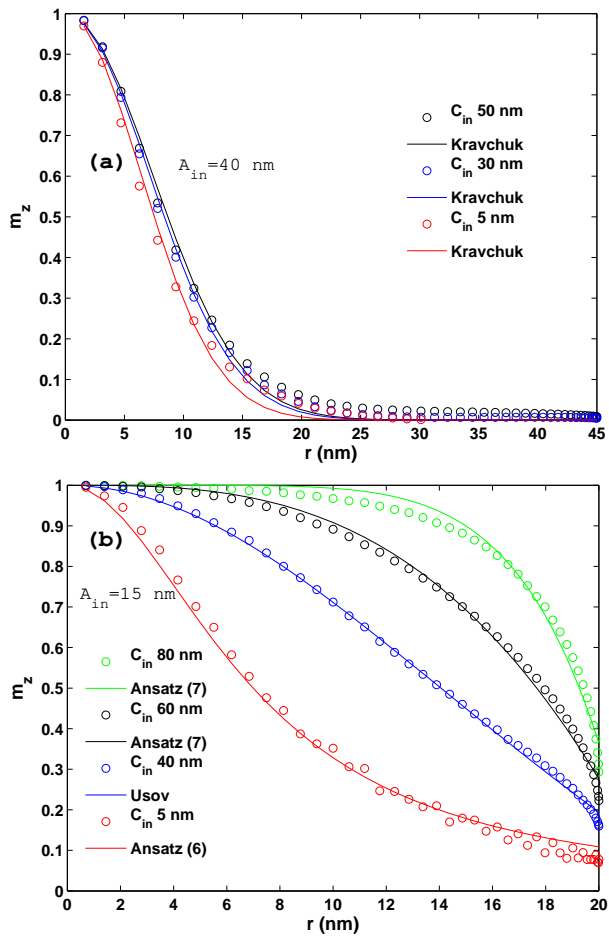}
\end{center}
\caption{The normalized z component of the magnetization as a function of $r$ for hemispheroidal shells (Thickness=10 nm). (a) Symbols correspond to the micromagnetic simulations for semi-principal axis $\rm A_{in}=40$ nm and different $\rm C_{in}$.  (b) Symbols indicate the micromagnetic simulations results for semi-principal axis $\rm A_{in}=15$ nm and three different different $\rm C_{in}$ values. Solid lines are the best ansatzes which fit to the simulations' results.} \label{profile_vortex10} 
\end{figure}

Figure $\ref{profile_vortex5}$(a) shows the profile of the normalized z component of the magnetization as a function of $r$, for $\rm A_{in}=40$ nm and shell thicknesses of 5 nm.  Kravchuk's ansatz (Eq. \ref{Kravchuk_eq}) can be a good approximation only for $\rm C_{in}/\rm A_{in} \leq$ 0.75. The deviation from the Kravchuk's ansatz increases with increasing the aspect ratio ($\rm C_{in}/\rm A_{in}$). The magnetization distributions for $1<\rm \frac{C_{in}}{A_{in}}<3$ do not fit at all to the Kravchuk's ansatz and do not correspond to Usov and Peschany's ansatz given by Eq. \ref{Usov_eq}. Therefore, to describe $m_z$ vs $r$ in this range we propose the one parameter ansatz:

\begin{equation}
\label{like_Lor_eq}
f(r)=\frac{\lambda^2}{\lambda^2+r^2}.
\end{equation} 
     
The profile of the normalized z component of the magnetization as a function of $r$, for $\rm A_{in}=15$ nm and shell thicknesses of 5 nm is plotted in Figure $\ref{profile_vortex5}$(b). Equation \ref{like_Lor_eq} can be a good approximation only for $\rm C_{in}/\rm A_{in} \leq$ 1.33. The magnetization distributions for $\rm \frac{C_{in}}{A_{in}}>2$ do not fit at all to the Kravchuk's ansatz or to Eq. \ref{like_Lor_eq} and do not correspond to Usov and Peschany's ansatz given by Eq. \ref{Usov_eq}. Therefore, to describe $m_z$ vs $r$ for $\rm \frac{C_{in}}{A_{in}}>2$ we suggest the two parameters ansatz:

 \begin{equation}
\label{Schultz_eq}
f(r)=1-(r/R_c)^{\alpha}.
\end{equation}
The fit is not good enough but it is the best simple ansatz.

The profile of the normalized z component of the magnetization as a function of $r$, for $\rm A_{in}=40$ nm and shell thicknesses of 10 nm is presented in Figure $\ref{profile_vortex10}$(a). The Kravchuk's ansatz is the best fit throughout the range ($\rm C_{in}/\rm A_{in} \leq$1.25).
 
Figure $\ref{profile_vortex10}$(b) depicts the profile of the normalized z component of the magnetization as a function of $r$, for $\rm A_{in}=15$ nm and shell thicknesses of 10 nm. The best fit for $\rm \frac{C_{in}}{A_{in}} \geq2.66$ is Eq. \ref{Schultz_eq}  and for $0.33<\rm \frac{C_{in}}{A_{in}}<0.66$ is Eq. \ref{like_Lor_eq}. Non of the above ansatz was good enough to describe $m_z$ vs $r$ for the middle size aspect ratios.

\section{Conclusions}

We present a detailed study of the ground state of magnetic nano hemispheroidal shells. In addition to the three magnetic ground states which exist in hemispherical magnetic shells we find another homogeneous ground state, the easy axis. This additional magnetic structure is the ground state only for small and elongated hemispheroidal shells. Like hemispherical shells \citep{SHEKA2013}, as the dimensions increase there is more preference for vortex state than the onion state. The vortex profile cannot be described by only one ansatz. We need to choose the right ansatz for each hemispheroidal shell. Start with Kravchuk's ansatz for the wide hemispheroidal shells with low aspect ratio ($\rm C_{in}/\rm A_{in}$), through Eq. \ref{like_Lor_eq} and end with the ansatz which is given by Eq. \ref{Schultz_eq} for the narrow hemispheroidal shells with large aspect ratio ($\rm C_{in}/\rm A_{in}$).  
 


\end{document}